\documentclass[aps,pra,amsmath,amssymb,floatfix,twocolumn,amsmath,superscriptaddress,twocolumn,nofootinbib,tighten,letterpaper]{revtex4-2}

\usepackage[colorlinks,linkcolor=blue,citecolor=blue,urlcolor=blue]{hyperref}

\usepackage{multirow}
\usepackage{subfigure}
\usepackage{color}
\usepackage{mathrsfs}
\usepackage{hyperref}
\usepackage[normalem]{ulem}
\usepackage{bm}

\usepackage{amssymb}   
\usepackage{amsmath}
\renewcommand\vec[1]{\ensuremath\boldsymbol{#1}} 

\usepackage{amsfonts, relsize, color}
\usepackage{graphics}
\usepackage{graphicx}
\usepackage{subfigure}
\usepackage{color}
\usepackage{comment}

\begin{document}
\title{Transport in strained graphene: Interplay of Abelian and axial magnetic fields}

\author{Aqeel Ahmed}
\thanks{These authors contributed equally to this work.}
\affiliation{Department of Physics, Lehigh University, Bethlehem, Pennsylvania, 18015, USA}
\affiliation{Department of Physics and Astronomy, Union College, Schenectady, New York, 12308, USA}

\author{Sanjib Kumar Das}
\thanks{These authors contributed equally to this work.}
\affiliation{Department of Physics, Lehigh University, Bethlehem, Pennsylvania, 18015, USA}

\author{Bitan Roy}
\affiliation{Department of Physics, Lehigh University, Bethlehem, Pennsylvania, 18015, USA}

\date{\today}

\begin{abstract}
Immersed in external magnetic fields ($B$), buckled graphene constitutes an ideal tabletop setup, manifesting a confluence of time-reversal symmetry (${\mathcal T}$) breaking Abelian ($B$) and ${\mathcal T}$-preserving strain-induced internal axial ($b$) magnetic fields. In such a system, here we numerically compute two-terminal conductance ($G$), and four- as well as six-terminal Hall conductivity ($\sigma_{xy}$) for spinless fermions. On a flat graphene ($b=0$), the $B$ field produces quantized plateaus at $G=\pm |\sigma_{xy}|=(2n+1) e^2/h$, where $n=0,1,2, \cdots$. The strain induced $b$ field lifts the two-fold valley degeneracy of higher Landau levels and leads to the formation of additional even-integer plateaus at $G=\pm |\sigma_{xy}|= (2,4,\cdots)e^2/h$, when $B>b$. While the same sequence of plateaus is observed for $G$ when $b>B$, the numerical computation of $\sigma_{xy}$ in Hall bar geometries in this regime becomes unstable. A plateau at $G=\sigma_{xy}=0$ always appears with the onset of a charge-density-wave order, causing a staggered pattern of fermionic density between two sublattices of the honeycomb lattice.           
\end{abstract}

\maketitle

\section{Introduction}

The interplay of Abelian ($B$) and axial ($b$) magnetic fields gives birth to a unique sequence of quantum Hall states and competing length scales of topological defect modes in two-dimensional Dirac materials~\cite{royBb1, royBb2, royBb3}. In this regard, a honeycomb membrane of carbon atoms, graphene, constitutes a tabletop platform where such a confluence can be experimentally studied. When buckled, the electromechanical coupling in a graphene flake produces time-reversal symmetric axial magnetic fields~\cite{crommie2010, manoharan2012, loh2012}. Just like its Abelian counterpart, uniform axial magnetic fields produce valley degenerate Landau levels (LLs). While the time-reversal symmetry (${\mathcal T}$) breaking external real magnetic field (also named here \emph{Abelian}) points in the same direction near two inequivalent valleys of the erstwhile hexagonal Brillouin zone (BZ), harboring massless Dirac fermions, its axial cousin points in the opposite direction near the complementary valleys. As the strain-induced internal magnetic field ($b$) couples two flavors of massless Dirac fermions residing near the opposite valleys with opposite signs~\cite{SM}, it is called \emph{axial}~\cite{Peskin2019-so} and it preserves the ${\mathcal T}$ symmetry. Therefore, the net effective magnetic fields near two valleys are of different magnitude in the simultaneous presence of Abelian and axial fields. Moreover, their directions depend on the relative strengths of these two fields. Despite tremendous experimental activities exploring the quantum Hall physics in graphene over the past several years~\cite{novoselov2005, pkim2005, ayoung2012, yacoby2012, elias2013, AFyoung2022}, the confluence of Abelian and axial magnetic fields has gained little experimental attention so far~\cite{linhe2015, Linhe2020}.

\begin{figure}[h!]
\includegraphics[width=1.00\columnwidth]{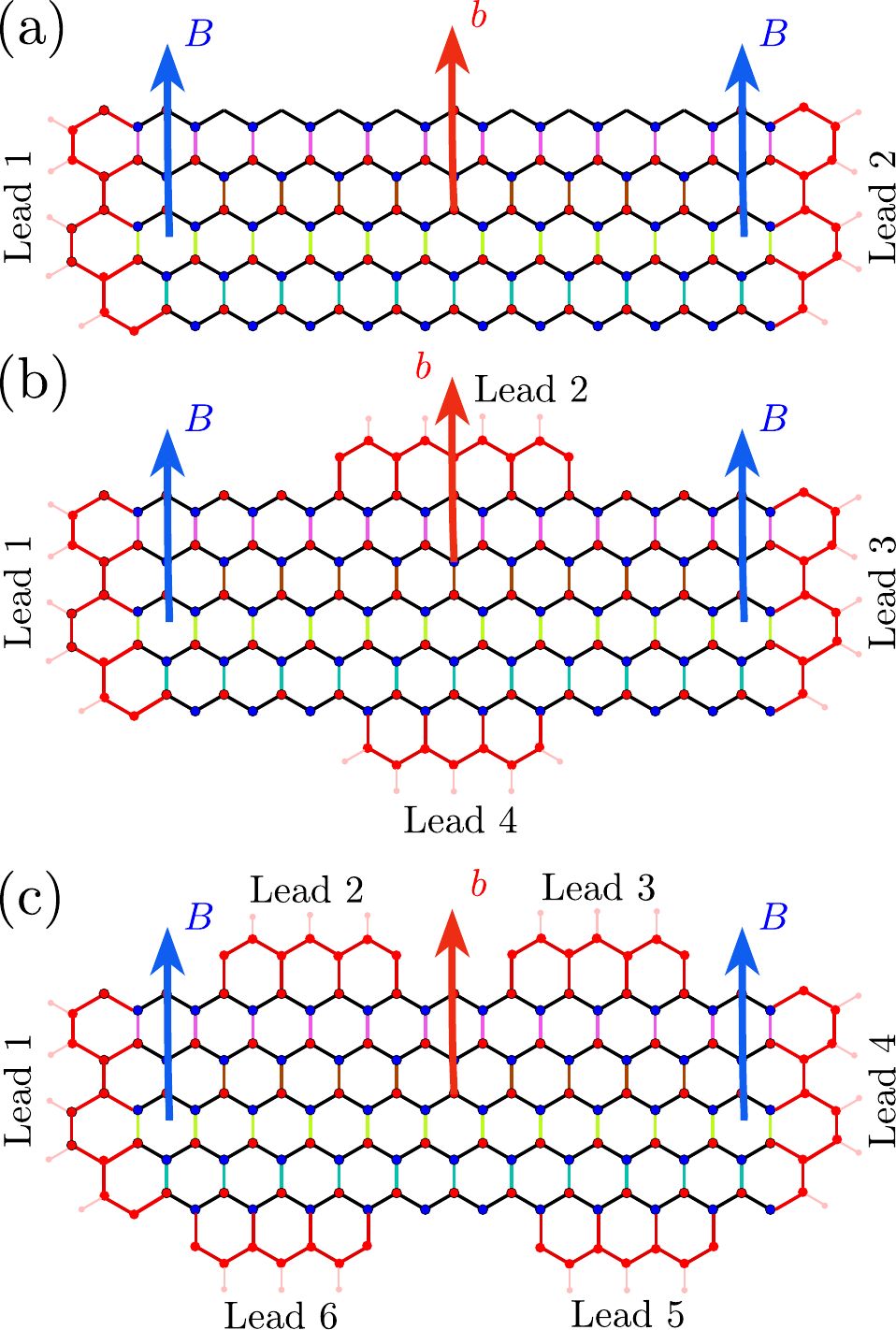}
\caption{Schematic (a) two- (b) four- and (c) six-terminal setups. Semi-infinite (red) hexagons are the leads, attached to the scattering region of length $L$ and width $W$, where the uniform hopping ($t$) is denoted by the black lines. The axial ($b$) [Abelian ($B$)] magnetic fields are shown by red [blue] arrows. Strain in graphene (yielding $b$) is produced by modified hopping $t[1+(\omega/W) y_k]$ along colored vertical bonds, where $y_k$ is the $y$ coordinate of the site living at the bottom of the corresponding colored bond, with $y_k \in [-W/2,W/2]$ and $y_k=0$ at the center of the scattering region in the $y$ direction, and $\omega$ sets the strength of the $b$ field [Eq.~\eqref{eq:strain}]. Sites from two sublattices are shown by red and blue filled circles.
}~\label{fig:schematic}
\end{figure}

Here we present a comprehensive numerical study of a mesoscopic graphene sample, subject to Abelian and/or axial magnetic fields in multiterminal arrangements using Kwant~\cite{Groth2014}. Specifically, we compute two-terminal conductance ($G$) and transverse Hall conductivity ($\sigma_{xy}$) in four- and six-terminal setups [Fig.~\ref{fig:schematic}]. Throughout, $G$ and $\sigma_{xy}$ are measured in units of $e^2/h$. As the competition between the $B$ and $b$ fields is insensitive to electronic spin (leaving aside the Zeeman coupling of the former), here we consider a collection of spinless fermions.

\begin{figure*}[t!]
\includegraphics[width=1.00\linewidth]{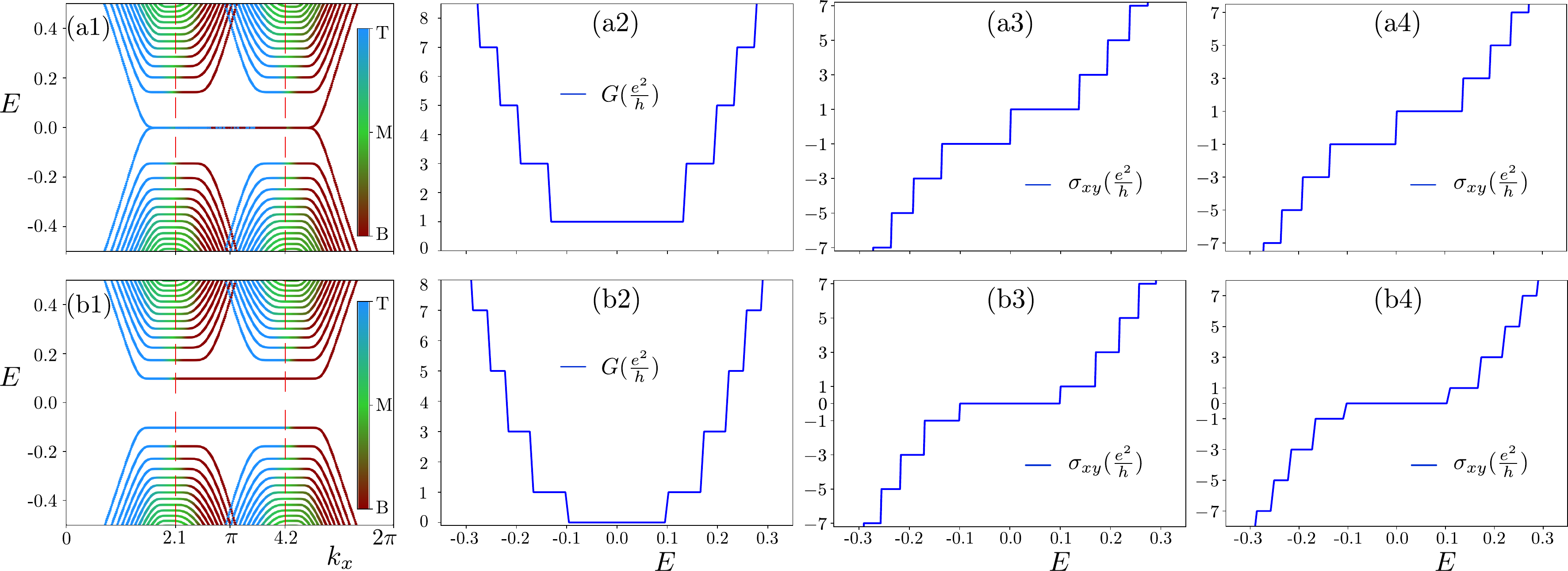}
\caption{Band structure of a zigzag graphene nanoribbon (lead Hamiltonian), containing $120$ sites in the $y$ direction and with translational invariance in the $x$ direction, for $t=1$, $B=2\times 10^{-3}$, $\omega=0.0$, and (a1) $\delta=0.0$ and (b1) $\delta=0.1$. The magnetic field induces flat twofold valley degenerate electronlike and holelike LLs with energies $\sim \sqrt{n}$, where $n\in\mathbb{Z}$ is the LL index. In (a1) and (b1), red dashed vertical lines indicate the locations of two Dirac points along the momentum axis. Modes that are living near the top (T) and bottom (B) edges (dispersive ones~\cite{SanjibGrapheneEdge2019}) and localized near the middle (M) of the system are color coded (see the color bars). Two-terminal conductance $G$, computed in a system of $L=W=120$, shows quantized plateaus at $2n +1$ in (a2) and (b2). The CDW order ($\delta$) gives birth to the $G=0$ conductance plateau in (b2), as it gaps out the zeroth LL. (a3) Four-terminal and (a4) six-terminal Hall conductivities in a $L=W=400$ system show quantized Hall plateaus at $\sigma_{xy}=\pm (2n+1)$  when $\delta=0$. A finite $\delta$ resolves the $\sigma_{xy}=0$ Hall plateau in (b3) four- and (b4) six-terminal measurements. Here, $G$ and $\sigma_{xy}$ are computed within the energy range $(-0.35, 0.35)$, containing $400$ grid points.
}~\label{fig:limita}
\end{figure*}

\section{Key results}

A flat graphene flake, subject to a real magnetic field ($B$), displays well-known quantized plateaus at $G=\pm |\sigma_{xy}|=2n+1$, manifesting the two-fold valley degeneracy of each LL, where $n=0,1,\cdots$~\cite{gysininsharapov2005} (Fig.~\ref{fig:limita}). A strain induced axial magnetic field ($b$) lifts the valley degeneracy of all the LLs, except the topologically protected zeroth one~\cite{royBb1, royBb2}. As a result when $B>b$, additional plateaus are observed at $G=\pm |\sigma_{xy}|=2,4,\cdots$ (Fig.~\ref{fig:limitb}). In the opposite limit when $b>B$, the higher LLs remain valley nondegenerate, and we find plateaus at $G=n+1$. Once the real magnetic field is switched off in an otherwise strained graphene, valley degenerate axial LLs lead to a plateau formation at $G=2n+1$, while $\sigma_{xy}=0$. See Fig.~\ref{fig:limitc}. However, when the axial field dominates over a finite $B$ field, the numerical procedure in four- and six-terminal geometries becomes unstable in Kwant, and we fail to capture any conclusive quantization of $\sigma_{xy}$. The formation of a charge-density-wave (CDW) order in all these cases gaps out the zeroth LL and in turn produces a $G=\sigma_{xy}=0$ plateau [Figs.~\ref{fig:limita}-\ref{fig:limitc}].

\section{Model}

The tight-binding Hamiltonian in graphene with only nearest-neighbor (NN) hopping ($t_{jk}$) reads as
\begin{equation}~\label{eq:grann}
H_0 = \big( -\sum_{\langle j,k \rangle} t_{jk} \; a_{j}^\dagger b_{k} + H.c. \big) 
+ \delta \big( \sum_{j} a_{j}^\dagger a_{j} - \sum_{k} b_{k}^\dagger b_{k} \big).
\end{equation}
The summation in the first term is restricted over three NN sites, the second term represents a sublattice resolved staggered potential (discussed below), and $a_{j}^\dagger$ and $a_{j}$ ($b_{j}^\dagger$ and $b_{j}$) are the fermionic creation and annihilation operators on $a$ ($b$) sublattices, respectively, constructed from the linear combinations of the Bravais vectors $\mathbf{a}_{1}=(1,0)d$ and $\mathbf{a}_{2}=(1,\sqrt{3})d/2$. Throughout, we set the lattice spacing $d=1$. A Fourier transformation of $H_0$ with $t_{jk}=t$ reveals linearly dispersing massless Dirac fermions near two inequivalent corners of the hexagonal BZ, suitably chosen at $\pm {\bf K}=2\pi (\sqrt{3},1)/(\sqrt{3}d)$~\cite{Semenoff1984}. The above model with only NN hopping ($\delta=0$) belongs to class BDI and the zigzag edge of graphene hosts localized zero-energy topological modes~\cite{Fujita1996}.

The orbital effect of an external $B$ field is incorporated via a Peierls substitution $t_{jk} \rightarrow t_{jk} \exp[2 \pi i\phi_{jk}]$~\cite{Peierls1933}. The flux phase $\phi_{jk}$ is given by the line integral $\phi_{jk} = \int_{j}^{k} \mathbf{A} \cdot \mathbf{dl}$ from site $j$ to site $k$. To introduce a uniform Abelian magnetic field ${\bf B}=B \hat{z}$, we choose a Landau gauge for the magnetic vector potential $\mathbf{A}=(-B y,0,0)$, such that ${\boldsymbol \nabla} \times \mathbf{A}={\bf B}$. It results in $t_{jk} \rightarrow t\exp[-i \pi \Phi(x_k-x_j)(y_j+y_k)]$, with $\Phi=Be \ell^{2}/h$ being the flux threading a unit cell of area $\ell^2$. Here $(x_j,y_j)$ is the real space coordinates of the site $j$ (see Supplemental Material~\cite{SM}). Application of an external magnetic field quenches the conical Dirac dispersion into a set of highly degenerate LLs, as shown in Fig.~\ref{fig:limita}(a1).

The axial or pseudo magnetic field in a graphene flake originates from a particular class of strain. For example, it can be modeled via a uniform modulation of one of the three NN bonds, here chosen to be the one perpendicular to the zigzag edge, from one end of the scattering region of width $W$ to the other~\cite{vishwanath2012, royherbutTIstrain, TianyuPhysRevResearch} (or by a Gaussian bump~\cite{PeetersJPhysCondMatt2017}). The hopping amplitude along such bonds between the $j$th and $k$th sites, respectively, located at $(x_j,y_j)$ and $(x_k,y_k)$ in the presence of Abelian and axial magnetic fields is~\cite{royBb2} 
\begin{equation}~\label{eq:strain}
t_{jk} \rightarrow t \left[ 1 + \omega \; \frac{y_k}{W} \right] \exp[-i \pi \Phi(x_k-x_j)(y_j + y_k)],
\end{equation}
where $y_k \in [-W,W]/2$, and the hopping along two other NN bonds retains its original strength $t$. Here, $\omega$ measures the strength of the strain, yielding an axial magnetic field $b \sim \omega$. Due to the strain gradient, the axial LLs acquire slightly inhomogeneous Fermi velocity, and hence they are not perfectly flat. Compare Fig.~\ref{fig:limitc}(a) and Fig.~\ref{fig:limita}(a1).

In half-filled graphene, the average fermionic density on any site is $1/2$ {\color{blue}when $\delta=0$}. Maintaining the overall filling unchanged, the system can develop a staggered pattern of fermionic density between two sublattices, resulting in a CDW order or staggered potential $\delta$ [Eq.~(\ref{eq:grann})]. In the presence of Abelian and/or axial magnetic fields, it can be supported by sufficiently weak NN Coulomb repulsion, following the spirit of \emph{magnetic catalysis}~\cite{miraskyetal2006, herbut2007, royherbut2011, roysau2014}. Here, however, we add such an order from the outset, yielding average fermionic densities $1/2 \pm \delta$ on $a$ and $b$ sublattices, respectively, as long as $\delta \leq 0.5$. The quantity $\delta$ measures the strength of the CDW order. Next we discuss two-, four-, and six-terminal setups to compute quantum transport in all these systems using Kwant~\cite{Groth2014}.

\subsection{Two-terminal transport}

~In a two-terminal setup, both the left and the right side of a square-shaped scattering region, made of graphene lattices, are connected to leads [Fig.~\ref{fig:schematic}(a)]. But we arrive at the same sequence of two-terminal conductance ($G$), and four- and six-terminal Hall conductivity ($\sigma_{xy}$) when computed with a rectangular scattering region, as shown schematically in Fig.~\ref{fig:schematic}. These symmetric leads are semi-infinite in the sense that they are connected to the system on one side and extend to infinity on the other side, preserving the translational invariance. In all our calculations, the leads and the scattering region have an identical Hamiltonian, as then the energy eigenstates in these two regions match at their interfaces, resulting in the smooth propagation of waves in the latter region, in turn stabilizing the numerical analyses~\cite{Groth2014}. Nonetheless, we verify that if the lead Hamiltonian corresponds to a pristine graphene Hamiltonian, our results do not change qualitatively. The leads are attached to the entire last set of sites of the scattering region. The corresponding unitary scattering matrix is given by 
\begin{equation}
S = \begin{pmatrix}
r & t'\\
t & r' 
\end{pmatrix},
\label{eq:smat}
\end{equation}
preserving the total probability of the incoming and outgoing modes. Here $r$ and $r'$ ($t$ and $t'$) are the reflection (transmission) parts of $S$. We compute the conductance $G =\text{Tr}(t^{\dagger}t)$ of the system from the transmission channels. The trace (`Tr') is taken over the conducting channels. Notice that there is no bias voltage in the system, and we compute $G$ in two-terminal setup, and $\sigma_{xy}$ in four- and six-terminal setups, as a function of varying energy $E$ that take into account the number of filled LLs and associated chiral edge modes below a certain energy $E$. Alternatively, one can compute $G$ and $\sigma_{xy}$ at fixed energy $E=0$ by varying the bias voltage in the scattering region, which leads to identical outcomes.

\begin{figure*}[t!]
\includegraphics[width=1.00\linewidth]{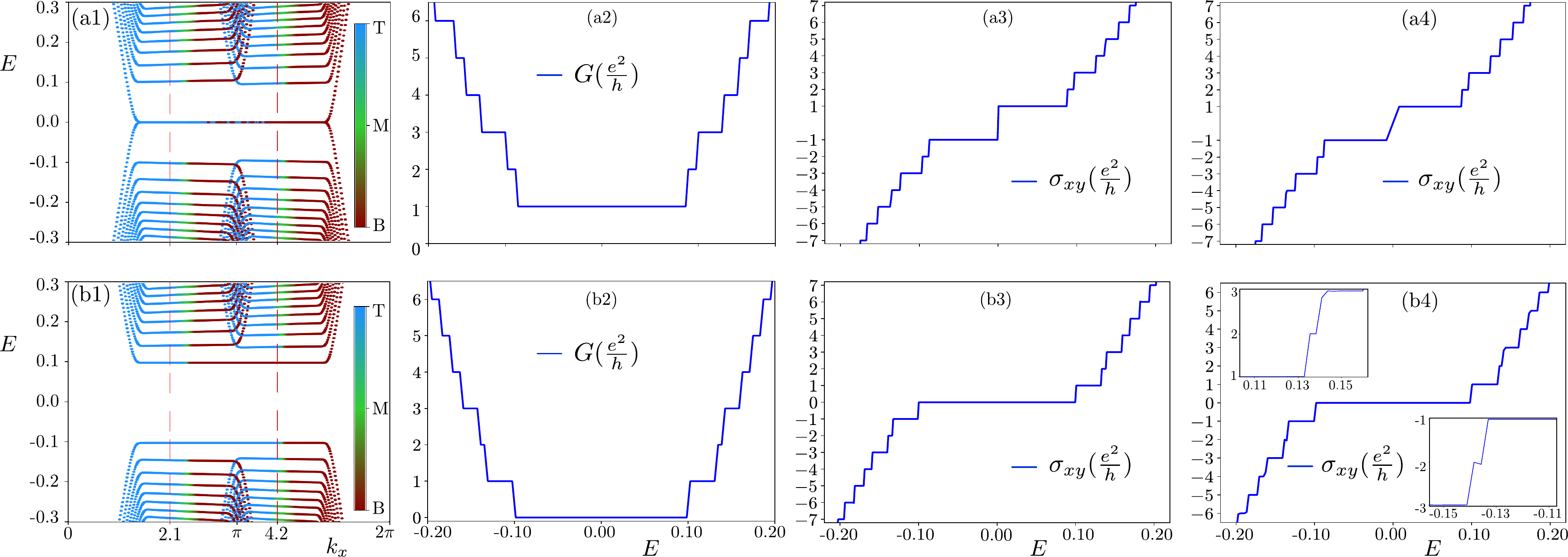}
\caption{Band structure of a zigzag graphene nanoribbon (lead Hamiltonian) containing $400$ sites in the $y$ direction and with translational invariance in the $x$ direction for $B=10^{-3}$, $\omega=0.2$, such that $B>b$, and (a1) $\delta=0.0$ and (b1) $\delta=0.1$, for which two-terminal conductance is shown in (a2) and (b2), respectively. The Hall conductivity $\sigma_{xy}$ in four- [six-] terminal setup is, respectively, shown in (a3) and (b3) [(a4) and (b4)]. Due to the valley degeneracy lifting of the higher LLs, additional even integer plateaus at $G=\pm |\sigma_{xy}|=2,4, \cdots$ are formed (see Fig.~\ref{fig:limita}), while the plateau at $G=\sigma_{xy}=0$ appears only in the presence of the CDW order. Insets in (b4) show the narrow plateaus for $\sigma_{xy}=\pm 2$. Numerical calculations are performed in a system with $L=W=600$, and with $400$ grid points in the energy window $(-0.25, 0.25)$. In (a1) and (b1), red dashed vertical lines indicate the locations of two Dirac points along the momentum axis. Modes that are living near the top (T) and bottom (B) edges (dispersive ones), and localized near the middle (M) of the system are color coded (see the color bars).
}~\label{fig:limitb}
\end{figure*}

\subsection{Four-terminal transport}

To capture the Hall response, one needs to go beyond the two-terminal arrangement, and consider a multi-terminal setup. Here we compute the four-terminal conductance in graphene, subject to real and/or pseudo magnetic fields. A current $j$ flows between Lead 1 and Lead 3, and the Hall voltage develops between the vertical Lead 2 and Lead 4, acting as the Hall probes [Fig.~\ref{fig:schematic}(b)]. Upon solving the current-voltage linear equation $\mathbf{j}=\mathbf{G} \mathbf{V}$, where $\mathbf{G}$ is the ${4\times 4}$ conductance matrix, we obtain the Hall voltage $V_{2}-V_{4}$, where $V_{p}$ is the voltage in the $p$th lead. The Hall conductivity in terms of $E_x=(V_3-V_1)/L$ and $E_y=(V_2-V_4)/W$, where $L$ ($W$) is the length (width) of the scattering region, is
\begin{equation}~\label{eq:Hall}
\sigma_{xy}=\frac{j_{x} E_{y}}{E_{x}^2+E_{y}^2}.
\end{equation}

\subsection{Six-terminal transport}

Since the six-terminal Hall bar geometry is most commonly employed in experiments to measure the Hall responses, here we also compute $\sigma_{xy}$ in this setup [Fig.~\ref{fig:schematic}(c)]. It allows us to compute the transverse Hall voltage between two vertical leads [Lead 2 (3) and Lead 6 (5)] and a longitudinal voltage between Lead 5 (3) and Lead 6 (2). We consider a current $j$ flowing only between Lead 1 and Lead 4 under the influence of an electric field $\mathbf{E}$. In the same spirit of the four-terminal calculation, here we have $E_x=(V_2-V_3)/L_{23}$ and $E_y=(V_3-V_5)/W$. Here, $L_{23}$ is the distance between Lead 2 and Lead 3, which we set to be $L/5$. Then $\sigma_{xy}$ can be computed from Eq.~(\ref{eq:Hall}). While computing $\sigma_{xy}$ in Kwant, it is important to employ a fine energy mesh to observe its sharp quantized plateaus.

\begin{figure}[t!]
\includegraphics[width=1.00\linewidth]{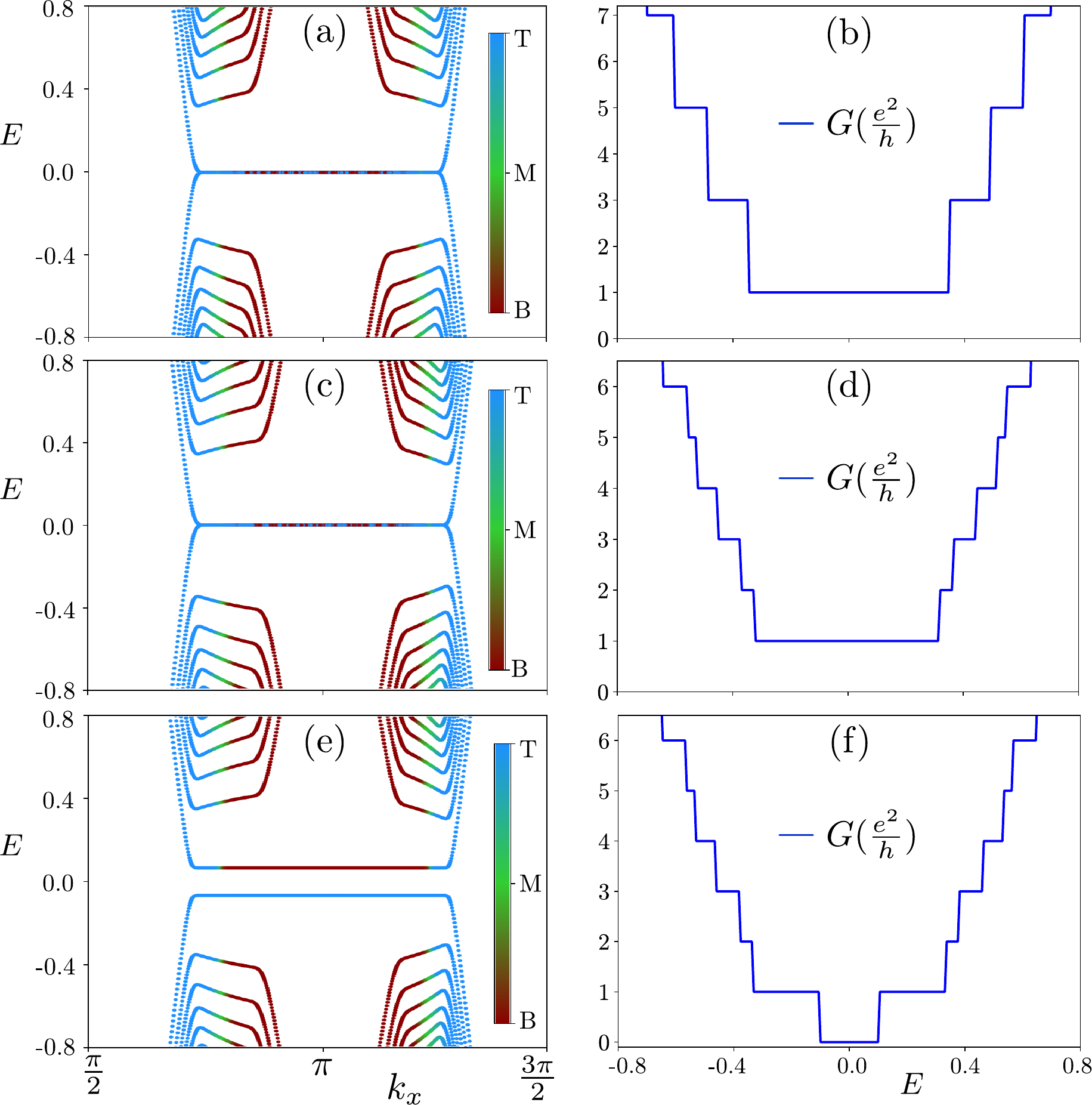}
\caption{Band structure of a zigzag graphene nanoribbon (lead Hamiltonian) with $400$ sites in the $y$ direction and translational invariance in the $x$ direction for $\omega=0.3$, $B=0$ [(a)] or $B=2 \times 10^{-5}$ [(c) and (e)], and $\delta=0.0$ [(a) and (c)] or $\delta=0.1$ [(e)], such that $b>B$ (always), presented over a part of the BZ containing well-separated LLs, realized by conveniently setting $t=10$. We compute two-terminal conductance ($G$) in a system with $L=W=400$. (b) With $B=0$, axial magnetic field ($b$) produces $G=2n+1$ plateaus, while additional even integer plateaus at $G=2,4, \cdots$ appear with a finite but weak $B$-field in (d). (f) The CDW stabilizes the $G=0$ plateau. In (a), (c) and (e), the Dirac points (see Figs.~\ref{fig:limita} and~\ref{fig:limitb}) fall outside the displayed region of $k_x$, and modes that are living near the top (T) and bottom (B) edges (dispersive ones), and localized near the middle (M) of the system are color coded (see color bars).  
}~\label{fig:limitc}
\end{figure}

\section{Results}.

To set the stage, we first consider a flat graphene, subject to Abelian magnetic fields ($B$). The system then supports twofold valley degenerate flat LLs, resulting from bulk cyclotron orbits, at energies $\pm \sqrt{2n B}$. The two-terminal conductance then shows monotonically increasing \emph{odd} integer quantized plateaus at $G=2n+1$, as the chemical potential is gradually tuned away from the half filling, thereby enhancing the number of occupied unidirectional quantized transmission channels. However, this setup is insensitive to the direction of the transmission channels and the nature of the carriers (electron or hole). These information unfold in four- and six-terminal setups, both featuring quantized Hall conductivity plateaus at $\sigma_{xy}=\pm (2n+1)$, respectively in the electron and hole doped regimes. The zeroth LLs near the opposite valleys live on complementary sublattices of graphene. Thus, formation of a CDW order gaps out the zeroth LL, thereby forming an insulator at half filling. Then an additional plateau at $G=\sigma_{xy}=0$ develops. These results are summarized in Fig.~\ref{fig:limita}.

Once buckled, the resulting axial magnetic field ($b$) lifts the valley degeneracy of all the LLs, as the effective magnetic fields are now ${\mathcal B}^{\pm}_{\rm eff}=(B \pm b)$ near the valleys at $\pm {\bf K}$, respectively. Two sets of particle-hole symmetric LLs then appear at energies $\pm [2 n |{\mathcal B}^{\pm}_{\rm eff}|]^{1/2}$ with the respective areal degeneracies $D_\pm=|{\mathcal B}^{\pm}_{\rm eff}|/(2 \pi)$. But, zeroth LLs remain pinned at zero energy, reflecting their topological protections~\cite{royBb1, royBb2, aharonovcaser}. Although it is challenging to extract $b$ directly in terms of $\omega$ from Eq.~(\ref{eq:strain}), notice that when $B=b$, only one valley with ${\mathcal B}^{+}_{\rm eff}=2B=2b$ hosts LLs, while the other one remains gapless as ${\mathcal B}^{-}_{\rm eff}=0$ therein. This is a quantum critical point, separating the field-dominated regime ($B>b$) from the strain dominated one ($b>B$). We first consider the former one(see Supplemental Material~\cite{SM}).

When $B>b$, the edge modes for two copies of nondegenerate LLs propagate in the same direction, as the effective magnetic fields ${\mathcal B}^{\pm}_{\rm eff}>0$ point in the same direction near two valleys. Consequently, the two-terminal conductance shows plateaus at all integers. The Hall conductivity in such a system can be computed from the St\v{r}eda formula $\sigma_{xy}=(\partial N/\partial B)_\mu$~\cite{PStreda1982}. Here, $N$ is the bulk electronic density and the derivative is taken at a fixed chemical potential ($\mu$). Under a small change of the magnetic field $\delta B$, the change in the number of states below (for electron doping) or above (for hole doping) the chemical potential is $\delta N= \Omega \left( n_+ + n_- \right) \delta B$, where $\Omega$ is the area of the graphene sample and $n_\pm$ is the number of filled LLs with areal degeneracies $D_\pm$, respectively, yielding $\sigma_{xy}=n_+ + n_-=n$. Therefore, in the field-dominated regime, the Hall conductivity only counts the number ($n$) of filled LLs at a fixed chemical potential, measured from the half filling. Concomitantly, we find $\sigma_{xy}=\pm (n+1)$ in both four- and six-terminal Hall bar geometries. In this regime, zeroth LLs near two valleys continue to reside on complementary sublattices~\cite{royBb1}. Thus, the formation of a CDW order resolves an additional plateau at $G=\sigma_{xy}=0$. These findings are displayed in Fig.~\ref{fig:limitb}.

Finally, we focus on the strain-dominated regime, as pseudo- or axial magnetic fields in buckled graphene can, in principle, be extremely large (a few hundred Tesla)~\cite{crommie2010, manoharan2012, loh2012}. When $B=0$, the system supports valley degenerate axial LLs, which are, however, slightly dispersive, possibly stemming from a spatially modulated Fermi velocity of Dirac fermions~\cite{FranzPhysRevB2020}. The edge modes, residing near the $\pm {\bf K}$ valleys, propagate in the opposite directions, manifesting the time-reversal symmetry, and are thus helical. In such a system, we find $G=2n+1$ in a two-terminal setup as it only counts the number of conducting edge modes, while being insensitive to their helicity. This observation strongly promotes the topological nature of the helical edge modes, leading to quantized transport (Fig.~\ref{fig:limitc}). But, $\sigma_{xy}=0$ in both four- and six-terminal arrangements due to the time-reversal symmetry (see Supplemental Material~\cite{SM}).

Application of a weak external magnetic field ($B<b$), lifts the valley degeneracy of all the axial LLs, except the zeroth ones. We then find $G=n+1$ (Fig.~\ref{fig:limitc}). In this regime, the St\v{r}eda formula implies $\sigma_{xy}=n_+ -n_-$, suggesting an \emph{oscillatory} behavior of the Hall conductivity as the chemical potential sweeps through the sea of electronlike or holelike LLs~\cite{royBb2}. Unfortunately, the Kwant-based numerical procedure in Hall bar geometries then becomes extremely unstable (see Supplemental Material~\cite{SM}), possibly due to its gauge dependence in multiterminal computation. And we fail to reach any conclusion on the quantization of $\sigma_{xy}$ in the strain dominated regime. When $b>B$, the zeroth LL wavefunctions are localized on one sublattice near both valleys in the bulk of the system, while they live on the complementary sublattice near its boundary. Therefore, a CDW order can be developed by creating a density imbalance between the bulk and the boundary of the system~\cite{roysau2014}, which gaps out the zeroth LL and in turn stabilizes a plateau at $G=0$ (Fig.~\ref{fig:limitc}).

\section{Summary and discussions}

Here we present a lattice-based extensive numerical analyses of quantum transport in strained graphene, immersed in external magnetic fields using Kwant~\cite{Groth2014}, in two-, four-, and six-terminal arrangements, capturing hallmarks of the interplay between Abelian ($B$) and axial ($b$) magnetic fields [Fig.~\ref{fig:schematic}]. Our findings are consistent with theoretical predictions from the continuum model in various limits, which include (a) flat graphene in ordinary magnetic fields ($b=0$) [Fig.~\ref{fig:limita}]~\cite{gysininsharapov2005}, (b) field-dominated regime ($B>b$) [Fig.~\ref{fig:limitb}]~\cite{royBb1}, and (c) strain-dominated regime (with only finite $b$ as well as $b>B$) [Fig.~\ref{fig:limitc}]~\cite{royBb2}. Possibly due to the gauge dependence in the numerical procedure in Kwant, we failed to underpin the expected oscillatory behavior of $\sigma_{xy}$ in the strain-dominated regime~\cite{royBb2}. Neither different gauge choices $\vec{A}=(0, B x, 0)$ and $\vec{A}=(- B y, Bx,0)/2$, nor the change in relative position of the voltage and current leads in comparison to the one shown in Fig.~\ref{fig:schematic} resolves this issue. To circumvent this limitation, in the future we will reinvestigate this problem using nonequilibrium Green's function method~\cite{NEGF1, NEGF2}. Furthermore, given that the thermal Hall conductivity ($\kappa_{xy}$) has been computed~\cite{ThermalHallGrapheneTh} and measured~\cite{ThermalHallGrapheneExp} in a flat graphene with $B$ fields, we will also compute quantized $\kappa_{xy}$, featuring the interplay of $B$ and $b$ fields. Our theoretical investigation should stimulate future experiments to showcase the intriguing confluence of magnetic fields in two-dimensional Dirac materials, which is still in its infancy~\cite{linhe2015, Linhe2020}.

\acknowledgments

S.K.D was supported by a Startup grant of B.R. from Lehigh University. A.A. was supported by the REU program through Grant No.\ NSF-PHY 1852010. We thank Suvayu Ali for technical support. B.R. was supported by NSF CAREER Grant No.\ DMR- 2238679.

\bibliography{ref}

\end{document}